\documentclass[pra,aps,twocolumn,groupedaddress,superscriptaddress]{revtex4-2}
\usepackage{graphicx}
\usepackage[nointegrals]{wasysym} %
\usepackage[export]{adjustbox}
\usepackage{amsmath,amsfonts,amssymb,latexsym}
\usepackage{hhline}
\usepackage{bm}
\usepackage{verbatim}
\usepackage{enumitem}
\usepackage{mathrsfs}
\usepackage{slashed}
\usepackage{empheq}
\usepackage[normalem]{ulem}
\usepackage{manfnt}
\usepackage{amsthm}

\usepackage{ragged2e}

\usepackage{caption}
\DeclareCaptionJustification{justified}{\justifying}

\newcommand{\p}{\partial}

\newcommand{\lan}{\langle}
\newcommand{\ran}{\rangle}

\newcommand{\ra}{\rightarrow}
\newcommand{\lra}{\leftrightarrow}

\newcommand{\wt}{\widetilde}

\newcommand{\bfzero}{{\mathbf{0}}}

\renewcommand{\(}{\left(}
\renewcommand{\)}{\right)}
\renewcommand{\[}{\left[}

\newcommand{\mt}{\mapsto}

\newcommand{\twp}{{2\pi}}

\newcommand\bpm            {\begin{pmatrix}}
	\newcommand\epm           {\end{pmatrix}}

\newcommand{\bs}{\bigskip}

\def\app#1#2{%
	\mathrel{%
		\setbox0=\hbox{$#1\sim$}%
		\setbox2=\hbox{%
			\rlap{\hbox{$#1\propto$}}%
			\lower1.1\ht0\box0%
		}%
		\raise0.25\ht2\box2%
	}%
}

\newcommand{\tw}{\textwidth}

\newcommand{\inv}{^{-1}}

\newcommand{\ope}\odot

\newcommand{\bi}{\begin{itemize}}
	\newcommand{\ei}{\end{itemize}}

\newtheorem{theorem}{Theorem}

\theoremstyle{definition}
\newtheorem{definition}{Definition}
\theoremstyle{definition}

\newcommand\bd            {\begin{definition}}
	\newcommand\ed            {\end{definition}}
\newcommand\bt            {\begin{theorem}}
	\newcommand\et            {\end{theorem}}
\newcommand\be            {\begin{equation}}
	\newcommand\ee            {\end{equation}}
\newcommand\ba            {\begin{aligned}}
	\newcommand\ea            {\end{aligned}}
\newcommand\bea{\begin{equation}\begin{aligned}}
		\newcommand\eea{\end{aligned}\end{equation}}

\usepackage{hyperref} 
\hypersetup{final}
\hypersetup{colorlinks, citecolor=red, linkcolor=red, urlcolor=red} 

\newcommand{\sss}{\subsubsection}
\renewcommand{\ss}{\subsection}

\renewcommand{\a}{\alpha}
\renewcommand{\b}{\beta}
\renewcommand{\d}{\delta}
\newcommand{\De}{\Delta}

\newcommand{\s}{\sigma}

\renewcommand{\o}{\omega}

\renewcommand{\r}{\rho}

\newcommand{\z}{\zeta}

\newcommand{\bfa}{\mathbf{a}}
\newcommand{\bfb}{\mathbf{b}}

\newcommand{\bfr}{\mathbf{r}}

\newcommand{\pp}{\mathbb{P}}
\newcommand{\rr}{\mathbb{R}}
\newcommand{\qq}{\qquad}

\newcommand{\zz}{\mathbb{Z}}

\usepackage[mathscr]{eucal} %

\newcommand{\req}{\rho^{\rm eq}}

\usepackage{braket}

\begin{document}
	\title{Scaling and localization in multipole-conserving diffusion}
	
	\author{Jung Hoon \surname{Han}}
	\email[Electronic address:$~~$]{hanjemme@gmail.com}
	\affiliation{Department of Physics, Sungkyunkwan University, Suwon 16419, South Korea}
	
	\author{Ethan Lake}
	\email[Electronic address:$~~$]{elake@mit.edu}
	\affiliation{Department of Physics, Massachusetts Institute of Technology, Cambridge, Massachusetts 02139, USA}
	
	\author{Sunghan Ro}
	\email[Electronic address:$~~$]{sunghan@mit.edu}
	
	\affiliation{Department of Physics, Massachusetts Institute of Technology, Cambridge, Massachusetts 02139, USA}
	
	\date{\today}
	\begin{abstract} 
		We study diffusion in systems of classical particles whose dynamics conserves the total center of mass. This conservation law leads to several interesting consequences. 
		In finite systems, it allows for equilibrium distributions that are exponentially localized near system boundaries. 
		It also yields an unusual approach to equilibrium, which in $d$ dimensions exhibits scaling with dynamical exponent $z = 4+d$. 
		Similar phenomena occur for dynamics that conserves higher moments of the density, which we systematically classify using a family of nonlinear  diffusion equations. 
		In the quantum setting, analogous fermionic systems are shown to form real-space Fermi surfaces, while bosonic versions display a real-space analog of Bose-Einstein condensation.
		
	\end{abstract}
	\maketitle
	
	\paragraph*{Introduction:}
	
	The phenomenon of diffusion is ubiquitous in physics, capturing the near-universal tendency of many-body systems 
	to relax at long times towards a uniform steady state. Diffusion is typically modeled using Fick's law by the equation $\p_t \r = \nabla \cdot (D \nabla \r)$, with $\rho$ the particle density and $D$ the diffusion constant. The assumptions going into the derivation of this equation are often very minimal, and the results are applicable to a broad range of physical phenomena. It is therefore important to understand situations in which conventional diffusive behavior breaks down. 
	
	An interesting question to ask in this direction is how diffusion is modified in systems with {\it constraints}, where restrictions are placed on the ways in which particles can move. A natural way of doing so is by the imposition of conservation laws that constrain particle motion. An example which has attracted much interest in the quantum dynamics \cite{pai2019localization,khemani2020localization,pollmann20,sala2020ergodicity,nandkishore21,feldmeier20,schulz2019stark,aidelsburger21,feng2022hilbert,moudgalya2021spectral} and many-body physics \cite{pretko17,prem18,pretko18,gromov20,glorioso22,gorantla22,pielawa11,sachdev02,lake1,lake2,stahl2022spontaneous,feldmeier,lake2023non,anakru2023non,chen2021fractonic,yuan2020fractonic} communities 
	is dynamics which conserves both the total particle number $N = \int dx\, \r(x,t)$ and the total dipole moment $Q_x = \int dx\, x \r(x,t)$, or equivalently the total center of mass $x_{\rm cm} = Q_x/N$ (working in 1d for simplicity). The requirement that both $N$ and $Q_x$ be time-independent is well-known \cite{pretko17} to mandate a continuity equation with two derivatives, viz.
	\begin{align} \partial_t \rho = - \partial_x^2 J.\label{eq:subdiffusive-eq} \end{align}  
	An expression for the current explored in recent hydrodynamically-motivated studies~\cite{nandkishore21,feldmeier20,gromov20,glorioso22,burchards22,glodkowski2023hydrodynamics} is
	\begin{equation} \label{eq:J_dipole_prev}
		J = \wt D \partial_x^2 \rho,
	\end{equation}
	with the coefficient $\wt D$ having dimensions of [length]$^4 \ $[time]$^{-1}$.
	This expression for $J$ then leads to subdiffusive behavior with dynamical exponent $z=4$. 
	
	\begin{figure}[b!]          
		\includegraphics[width=.98\columnwidth]{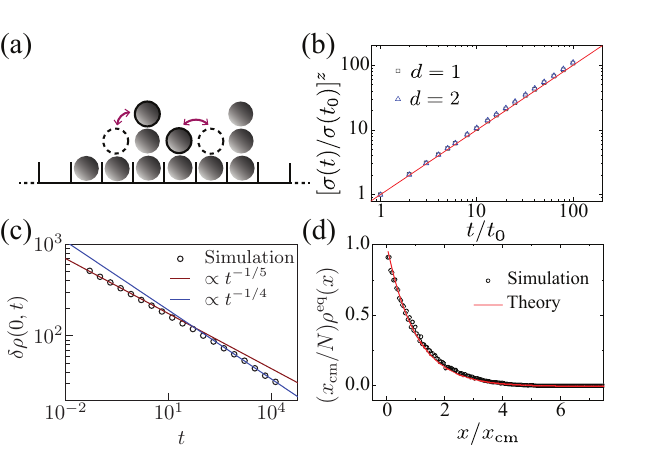}
		\vspace{1em}
		\caption{\label{fig:dipole} (a) Schematic of our dipole-conserving hopping process, where particles hop toward opposite directions in pairs. (b) Evolution of the standard deviation of the density $\sigma(t)$ starting from initial Gaussian distribution. In both 1d (circles) and 2d (triangles), $\s(t)$ exhibits an excellent scaling collapse with dynamical exponent $z=4+d$ (the straight line has slope 1, demonstrating $\sigma(t) \propto t^{1/z}$). (c) Relaxation of a large number $N_0=6000$ of particles localized at $x=0$ in the presence of a small constant-density background $\rho_0=10$, showing a crossover from $z=5$ at short times to $z=4$ at long times.
			(d) Exponentially-localized equilibrium density profile for 1d dipole-conserving diffusion on the half-space $x>0$.}
		\label{fig:1} 
	\end{figure}
	
	In this study, we show that Eq.~\eqref{eq:J_dipole_prev} only describes small fluctuations about a nonzero background density, and the full physics of dipole-conserving diffusion is much richer. By deriving an explicit lattice master equation for a natural class of dipole-conserving dynamics (see 
	Fig.~\ref{fig:dipole}(a)), we find a current which instead has the nonlinear form
	\begin{equation} \label{eq:J_dipole_our}
		J = D [ \rho \partial_x^2 \rho - (\partial_x \rho)^2 ] = D \r^2 \p_x^2 \ln (\r) ,
	\end{equation}
	with $D$ having dimensions of [length]$^5\ $[time]$^{-1}$ \footnote{Note that ordinary diffusion equation can be rewritten in a similar form as $\partial_t \rho = -\partial_x J$ with $J = -D \rho \partial_x \ln (\rho)$.}. 
	
	A scaling analysis of \eqref{eq:J_dipole_our} yields a dynamic exponent of $z = d+4$ in $d$ dimensions, with the dimension dependence arising from the fact that $J$ is a nonlinear function of $\rho$. 
	This scaling governs the relaxation 
	that occurs around a zero-density background. 
	On the other hand, the relaxation of {\it small} fluctuations around a {\it nonzero} background density $\bar \rho$ evolves with $z=4$, as can be seen by noting that \eqref{eq:J_dipole_our} reduces to \eqref{eq:J_dipole_prev} with respect to the density difference $\delta \rho \equiv \rho - \bar{\rho}$ after linearizing in $\delta\rho$. 
	A further notable feature of the expression \eqref{eq:J_dipole_our} is that $\r(x) \propto e^{-x/\ell}$ is a steady-state solution for all $\ell$ (see Fig.~\ref{fig:dipole}(d)). This reflects the principle of entropy maximization, suggesting that the form of \eqref{eq:J_dipole_our} should be general regardless of the specifics of the underlying microscopic dynamics.

	Our analysis extends to dynamics conserving higher multipole moments of density. As an example, in quadrupole-conserving dynamics that preserves both $x_\mathrm{cm}$ and the standard deviation $\sigma \equiv (N^{-1} \int dx\, x^2 \rho(x,t))^{1/2}$, the resulting steady-state distributions are Gaussians, with the conservation of $\sigma$ preventing the spreading of the density field.
	
	Finally, the principles identified in our study are also applicable to particles obeying quantum statistics. Multipole-conserving fermions turn out to form real-space Fermi surfaces, while bosons exhibit a real-space analog of Bose-Einstein condensation, with a macroscopic number of bosons condensing at a single site. 
	
	\paragraph*{Steady-state distribution:} We begin by describing the origin of the exponentially-localized steady-state shown in Fig.~\ref{fig:dipole}(d). In the absence of the potential energy due to interactions, the steady-state density $\req(x)$ 
	can be derived by maximizing the entropy $S = - \int dx\, \rho (x) \ln \rho (x)$ subject to the conservation of the particle number $N = \int dx\, \rho(x)$ and the center of mass $x_\mathrm{cm} = N^{-1} \int dx \, x \rho(x)$. Carrying out the extremization using Lagrange multipliers directly yields an exponentially-localized density profile. In particular, if the particles are confined to the half-line $x >0$, the equilibrium density is 
	\begin{equation} 
		\label{reqcm}
		\req (x) = \frac{N}{x_\mathrm{cm}} e^{-x / x_\mathrm{cm}}.
	\end{equation}
	The distribution of dipole-conserving particles is thus analogous to the Boltzmann distribution in the canonical ensemble, with $x_\mathrm{cm}$ playing the role of an effective temperature controlling the real-space width of the particle distribution. 
	
	It is important to note that this approach of maximizing $S$ tacitly assumes ergodic dynamics, an assumption that can break down in certain conditions. As an extreme example, a single isolated particle is immobile due to the dipole constraint, leading to completely non-ergodic behavior. More broadly, at low particle densities, the dynamics can be non-ergodic when only the strictly local hopping processes are considered \cite{khemani2020localization,pollmann20,sala2020ergodicity}.
	In what follows we will avoid this issue by always working with a large number of particles per site, far above \cite{skinner23,morningstar20} the critical density for the onset of non-ergodicity. 
	
	\paragraph*{Derivation of the diffusion equation:}
	
	We now consider a microscopic model within which \eqref{eq:J_dipole_our} can be derived explicitly. To proceed, consider a continuous-time lattice gas where a pair of particles close to each other are randomly chosen and then displaced locally in a way that conserves $x_\mathrm{cm}$. Denoting the number of particles at site $x$ as $\r_x$, the particles can ``diffuse out" as
	\begin{equation}
		(\Delta \r_x, \Delta \r_{x+1}, \Delta \r_{x+2}, \Delta \r_{x+3}) = (+1, -1, -1, +1) \label{eq:diffuse-out} 
	\end{equation}
	with a rate taken to be $r dt \r_{x+1} \r_{x+2}$ (here $\Delta \r$ gives the change in $\r_x$ between $t+dt$ and $t$). Similarly, the particles may ``diffuse in" as
	\begin{equation}
		(\Delta \r_x, \Delta \r_{x+1}, \Delta \r_{x+2}, \Delta \r_{x+3}) = (-1, +1, +1, -1) \label{eq:diffuse-in} 
	\end{equation}
	with a rate $r dt \r_x \r_{x+3}$ (these rates can be readily checked to satisfy detailed balance \cite{si}). 
	The fact that these rates are density-dependent is crucial for obtaining the nonlinear diffusion equation to follow. This should be contrasted with a distinct class of models inspired by quantum circuits \cite{pai2019localization,khemani2020localization,pollmann20,sala2020ergodicity,nandkishore21,feldmeier20,schulz2019stark,aidelsburger21,feng2022hilbert,moudgalya2021spectral}, where the dynamics is driven by randomly implementing dipole-conserving hops at a density-{\it in}dependent rate.

	Using the rates computed above, we obtain a master equation for the evolution of $\r_x$ of the form
	\begin{align}
		\label{dipmaster}
		\nonumber \partial_t \r_x &= r (\r_{x-1} \r_{x+2} + \r_{x-2} \r_{x+1} + \r_{x+1} \r_{x+2} + \r_{x-2} \r_{x-1}) \\
		& - r \r_x (\r_{x-1} + \r_{x+1} + \r_{x-3} + \r_{x+3}) . 
	\end{align}
	We now perform a derivative expansion of the above equation, yielding
	\begin{equation}
		\partial_t \rho = - D \partial_x^2 \left[  \rho \partial_x^2\rho - (\partial_x \rho)^2 + (\cdots)\right], \label{eq:dipolar-diff-eq} 
	\end{equation}
	where we have defined $D \equiv 4r a^4$ with $a$ the lattice spacing, and where $(\cdots)$ indicates terms with higher orders of derivatives. For distributions with a characteristic length scale $\ell$ (e.g. $\ell = x_{\rm cm}$ in \eqref{reqcm}), these terms are suppressed in powers of $a/\ell$, and \eqref{eq:J_dipole_our} is then reproduced in the limit $a / \ell \ll 1$. More details on the validity of this limit, as well as what happens when different types of elementary hopping processes are considered, can be found in the SI \cite{si}. The nonlinear dependence of $J$ on $\r$ is a simple consequence of the fact that dipole-conserving motion always involves {\it pairs} of particles.

	The unusual scaling observed in the relaxation towards equilibrium follows from the nonlinearity of the new diffusion equation \eqref{eq:dipolar-diff-eq}.  Indeed, the fact that $\r$ has dimensions of inverse length immediately yields a dynamic exponent of 
	$z=4+1$, which we confirm in numerics by measuring the scaling of the standard deviation $\sigma$ of the density evolving from a Gaussian initial state (Fig.~\ref{fig:dipole}(b)).
	
	Nevertheless, {\it small} fluctuations around a nonzero background density adhere to a dynamic scaling with $z=4$, aligning with the hydrodynamic expectation \eqref{eq:J_dipole_prev}. This is seen by setting $\r(x,t) = \req(x) + \d\r(x,t)$, linearizing in $\d\r$ given that $\d\r \ll \req(x)$, and dropping all derivatives of $\req(x)$, under which \eqref{eq:J_dipole_our} reduces to  
	\be \label{linearized} \p_t \d\r \approx -D\req(x)\p_x^4 \d\r,\ee 
	with $\wt D \equiv D \req(x)$ constituting an effective (sub)diffusion constant. This means that a quench from a strongly inhomogeneous density distribution will exhibit a dynamical crossover, with $\d \r(t) \sim t^{-1/5}$ at short times (where the fluctuations about the background are large) switching to $\d\r(t)\sim t^{-1/4}$ at long times (where fluctuations are small). It is especially remarkable that the density relaxes {\it faster} at longer times, since in a situation where multiple modes with distinct $z$ contribute to the relaxation, one would normally expect the long-time behavior to be dominated by the {\it slower} modes. 
	To verify this crossover numerically, we consider an initial configuration where a localized peak of $N_0$ particles are placed on top of a uniform background of density $\rho_0$ (see Fig.~\ref{fig:dipole}(c)).   
	Initially the background density is negligible in comparison to the localized peak, 
	and the relaxation of $\delta \rho$ at the peak center follows $\delta \rho(0,t) 
	\propto t^{-1/5}$. However, at later times when $\delta \rho$ becomes comparable to $\rho_0$, the scaling exhibits dynamic exponent showing a clear crossover to 
	$\delta \rho(0,t) \propto t^{-1/4}$~\footnote{The results of direct scaling of density fields with the dynamic exponents are also presented in SI~\cite{si}}. 
	
	We briefly note two extensions of this analysis. The first is to incorporate the effect of noise into the deterministic equation
	\eqref{eq:J_dipole_our}. An application of standard techniques \cite{lefevre07,hexner17} to the present dipole-conserving case gives the Langevin equation \cite{si} (the structure of which can also be inferred on symmetry grounds alone \cite{guo2022fracton})
	\be \label{dip_langevin} \partial_t \rho (x, t) = -D\partial_x^2 (\r^2\p_x^2 \ln(\r)) + \sqrt{2D} \p_x^2  \eta (x, t),
	\ee 
	where the noise field $\eta$ satisfies the fluctuation-dissipation theorem (FDT),
	\begin{align} \langle \eta (x, t) \eta(x', t') \rangle = \delta (t - t') \delta (x - x') \r^2(x,t). \label{eq:FDT} \end{align} 
	relating the strength of the noise on the l.h.s. to the ``mobility", which as shown in~\eqref{eq:J_dipole_our} is proportional to $\rho^2$ \footnote{The FDT holds for multipole-constrained diffusive system as well, being as it is applicable to any near-equilibrium situation.}. Linearizing the Langevin equation in $\d\r$ allows us to derive the structure factor \cite{si}
	\be S(k) \sim \lan \d\r(k,t) \d\r(-k,t)\ran = \req, \ee 
	which is interestingly the same result as for conventional diffusion.

	The second extension is the generalization of \eqref{eq:J_dipole_our} to $d$ dimensions, where the dipole-constrained diffusion equation reads $\p_t \r = \p_a \p_b J^{ab}$, with the current 
	\be J^{ab} =-D \r^2 \p^a \p^b \ln(\r).\ee 
	The resulting dynamical exponent is then
	\be z=4+d. \ee  
	We verify this numerically by repeating the scaling analysis for a system of dipole-conserving particles on a 2d square lattice, with the results of Fig.~\ref{fig:1}(b) indeed demonstrating excellent $z=6$ scaling.
	
	\paragraph*{Quadrupole conserving diffusion: } 
	
	\begin{figure}
		\includegraphics[width=.98\columnwidth]{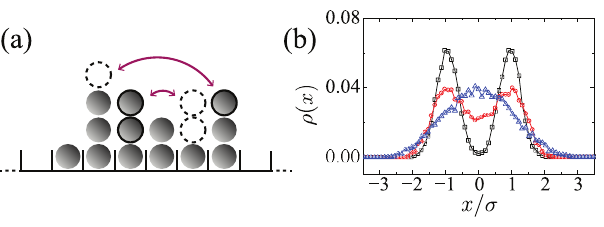}
		\vspace{1em}
		\caption{(a) A schematic illustration of a 1d quadrupole conserving diffusion process: two particles move two sites in one direction, while another particle moves three sites in the opposite direction. (b) The time evolution for quadrupole-conserving diffusion.  
			The density is initially given as two widely separated Gaussian packets, which merge over time (black $\to$ red $\to$ blue) into a single Gaussian profile, with the standard deviation $\sigma$ remaining conserved (solid curves are guides to the eye)}.
		\label{fig:2}
	\end{figure}
	
	We now move on to considering dynamics that conserves the quadrupole moment $Q_{xx} = \sum_x  x^2 \r_x$ (related to the standard derivation as $\s = \sqrt{Q_{xx}/N}$) in addition to the center of mass. Any quadrupole-conserving processes must involve the coordinated motion of at least three particles \cite{si}, with the simplest allowed three-particle process given by  $(\Delta \r_x, \Delta \r_{x+1}, \cdots, \Delta \r_{x+4}) = (\pm 1, \mp 2, 0, \pm 2, \mp1)$, as depicted in Fig.~\ref{fig:2}(a).
	By writing down a master equation via manipulations similar to those that led to \eqref{dipmaster}, we obtain the quadrupole-conserving diffusion equation
	\be \label{quaddiff} \p_t \r = \p_x^3 J,\qq J  = D \r^3 \p_x^3 \ln(\r).\ee
	
	Maximizing the entropy subject to the conservation of both dipole and quadrupole moments yields Gaussian steady states of the form
	\be \req(x) = \frac N{\sqrt{\twp \s^2}} e^{-(x-x_{\rm cm})^2 / 2\s^2},\ee 
	which are also static zero-current solutions to \eqref{quaddiff}. 
	This thus predicts the scenario whereby a generic initial distribution eventually ``congeals'' into a Gaussian, and then remains that way for all further times. The simulation results shown in Fig. \ref{fig:2}(b) indeed confirm this expectation. 
	
	Due to the presence of the nonzero conserved length scale $\s$, %
	no single-parameter scaling ansatz $\r(x^z/t)$ with finite $z$ is possible in the present case. Small deviations about $\req$ will however relax with $z=6$, as seen by linearizing \eqref{quaddiff}. 
	
	\paragraph*{General multipole moments: } We further generalize the discussion to a $d$-dimensional system that conserves $ \int_\bfr p_n(\bfr) \r_\bfr$ for all degree-$n$ polynomials $p_n(\bfr)$. Maximizing the entropy yields equilibrium states formed as exponentials of degree-$n$ polynomials in the coordinates, which leads to generalizations of \eqref{eq:subdiffusive-eq}, \eqref{eq:J_dipole_our} and \eqref{quaddiff} as 
	\be \p_t \r = - \p_A J^A,\qq J^A = (-1)^{n+1}D\r^{m} \p^A \ln \r , \label{eq:most-general-relation} \ee 
	where $A = \{a_1,\dots,a_{n+1}\}$ is a composite index and $\p_A \equiv \p_{a_1} \cdots \p_{a_{n+1}}$. The minimum value for $m$ in the definition of the current $J^A$ (which determines the mobility $D\r^m$) is equal to the minimum number of particles needed to form a microscopic $n$-pole conserving hopping process. In \cite{si} we show that $m=n+1$ for $n\leq 10$ and $n=12$; determining whether or not $m=n+1$ for all $n$ turns out to be equivalent to a longstanding open problem in number theory \cite{borwein2002computational}. As a further consistency check, the two relations in (\ref{eq:most-general-relation}) imply that the entropy $S = - \int \r \ln \r$ is monotonically increasing with time, 
	as they yield
	\begin{align}
		\frac{dS}{dt}  = D \int dx \rho^m \left(  \partial^A [ \ln \rho + 1] \right)^2  \geq 0 . 
	\end{align}

	\paragraph*{Quantum mutipolar diffusion: } 
	
	So far, we have treated the objects undergoing diffusion as classical (distinguishable) particles. We now generalize by letting the particles obey quantum statistics. This is particularly important in light of the fact that a natural experimental realization of multipole-conserving dynamics is found in quantum particles in strongly tilted optical lattices \cite{aidelsburger21,bakr20,weitenberg22}. 
	
	We begin by generalizing the expression~\eqref{eq:J_dipole_our} obtained for dipole-conserving dynamics. This can be done by suitably modifying the lattice master equation. In the classical case, we took the rate for a pair at locations $x_1,x_2$ to hop to $x_3,x_4$ to be proportional to $\r_{x_1} \r_{x_2}$. To generalize, we simply modify this to $\r_{x_1} \r_{x_2} (1+\z \r_{x_3})(1+ \z \r_{x_4})$, where $\z=+1,-1$ for bosons and fermions, respectively. These factors account for the ``statistical interactions'' (attractive or repulsive) experienced by indistinguishable particles, and follow from demanding that the dynamics satisfy detailed balance~\cite{si}. 
	
	With these modifications to the hopping rates, the same approach used previously produces~\cite{si}
	\bea \label{quantum_current_main} J & = D (\r \p_x^2 \r - (\p_x\r)^2 + \z (\r^2 \p_x^2 \r - 2 \r (\p_x\r)^2 )) \\ 
	& = D\r^2 (1+\z\r)^2 \p_x^2 \ln \( \frac{1}{1/\r+\z}\).\eea
	From the second line, it is easy to see that the zero-current steady states are real-space Bose-Einstein ($\z=1$) or Fermi-Dirac ($\z=-1$) distributions: 
	\be
	\req(x) = [e^{(x-\mu)/T}-\z]\inv,\ee 
	as could be expected on entropy-maximization grounds (here $T,\mu$ are determined by $N,Q^x$). Note that~\eqref{quantum_current_main} reduces to~\eqref{eq:J_dipole_our} in the (classical) low-density limit ($\z=0$). However, for bosons at high densities, $J$ becomes {\it cubic} in $\r$, giving a crossover from $z=5$ to $z=6$ dynamics as the density is increased.  
	
	We now point out a few curious features of the steady-state distributions that arise in this case. 
	For illustration, first consider fermions in a 1d box $x \in [-L/2,L/2]$. At half-filling $N = L/2$ one finds $\mu=0$, with $T$ fixed by the center of mass: $T = \infty$ corresponds to $x_{\rm cm} = 0$, while $T = 0^\pm$ gives $x_{\rm cm} = \mp L/4$ (here ${\rm sgn}(T) = -{\rm sgn}(x_{\rm cm})$). The fermions thus form a Fermi surface in {\it real space}, with the Fermi ``radius'' roughly fixed by $N$, and the ``temperature'' roughly fixed by $x_{\rm cm}$ \footnote{
		Since the dynamics freezes out below a critical (hopping-range-dependent) density, we expect that the density will follow a Fermi-Dirac distribution only in regions where $\req(x) \gtrsim 1/\ell_{max}$, where $\ell_{max}$ is the range of the longest-ranged hopping process \cite{morningstar20,skinner23}.}.

	Now consider $N$ dipole-conserving bosons in the positive-coordinate space $\pp^d \equiv [0,\infty)^d$, and suppose for simplicity that all components of the dipole moment $Q^a$ are equal. 
	The new physics afforded by bosonic statistics is the possibility of forming a Bose-Einstein condensate (BEC). The BECs that form in the present case are distinguished by the fact that the ordering occurs in real space, with a macroscopic number of bosons condensing onto a single lattice site. 
	
	The transition ``temperature'' $T_*$ at which a real-space BEC forms is determined by the point at which the chemical potential vanishes, calculated by solving
	\be\label{beccond} N = \int_{\pp^d} \frac{d^dr}{e^{ \sum_a r_a / T_*} - 1}\ee 
	for $T_*$. For $d=1$ this integral has a {\it short}-distance divergence, which precludes ordering by mandating that $T_* \ra 0$ (cf. the {\it long}-distance divergences that destroy conventional 1d BECs). When $d>1$, a BEC can form at $T_* \propto N^{1/d}$~\cite{si}. By calculating the dipole moment $Q^x$ at $T=T_*$, we find that a BEC forms when $N > N_*$, with 
	\be N_* \propto  (Q^x/a)^{\frac d{d+1}},\ee 
	where 
	we have restored the lattice constant $a$. To understand this, imagine increasing the number of particles in the system while keeping $Q^x$ fixed. For $N < N_*$, the system will smoothly adjust its overall density profile as particles are added. When $N>N_*$, all of the particles subsequently added to the system will join the condensate at $\bfr=\bfzero$, where they do not contribute to the dipole moment. In this regime the system will thus possess a condensate at the origin, together with a surrounding ``normal fluid'' localized nearby. 
	
	Bosons with higher moment conservation behave similarly. As an example,  quadrupole-conserving bosons in infinite space have $\req(r) = [e^{ (r^2-\mu)/T } - 1]\inv$ (assuming rotational symmetry), rendering the problem equivalent to the nonrelativistic ideal Bose gas after the interchange of coordinates and momenta. A short-distance divergence prevents a BEC from forming when $d<3$ \cite{si}. For $d\geq 3$, the critical boson number is determined in terms of the quadrupole moment $Q^{xx}$ as $N_* \propto (Q^{xx}/a^2)^{\frac{d}{d+2}}$.
	Understanding other physical properties of these unusual BECs, especially the nature of their condensation transitions, are questions that we leave to future work. 
	
	\paragraph*{Discussion:}
	
	We have demonstrated that multipole-conservation laws can radically alter how diffusion occurs. Even in the absence of interactions, these conservation laws lead to exponentially localized steady-states, and unconventional approaches to equilibrium. In the case of dipole conservation, we have explicitly demonstrated that relaxation occurs with an anomalously large scale-dependent dynamical exponent, with large-scale features relaxing with $z=4+d$, and short-scale ones with $z=4$. 
	
	It would be of great interest to pursue experiments realizing multipole-conserving dynamics in the lab. A natural platform in this regard is found in strongly tilted optical lattices \cite{khemani2020localization,sala2020ergodicity,aidelsburger21,bakr20,weitenberg22}, where the lattice tilt enforces emergent dipole conservation over a long prethermal time scale. The breakdown of diffusion due to ergodicity breaking can be avoided by working at sufficiently high densities, with the envisioned experiment rather similar to the one performed in \cite{weitenberg22}. As an example, we predict that sufficiently high-density and sufficiently weakly-interacting bosons in a strongly-tilted 1d lattice will relax with a dynamical exponent $z = 6$, and will admit equilibrium distributions that condense on system boundaries. For fermions, the equilibrium density profile will instead follow the Fermi-Dirac distribution.

	There are many natural extensions of our work. Instead of full multipole moment conservation, one could examine {\it non-maximal} multipole groups~\cite{gromov2019towards,stahl2022spontaneous} in $d>1$, where the dynamics conserves $\sum_\bfr P(\bfr) \r_\bfr$ for only a subset of degree-$n$ polynomials $P(\bfr)$, {\it e.g.} a 2d system conserving $P(\bfr) = x^2, y^2$, but not $P(\bfr) = xy$. 
	It may also be interesting to examine the importance of the lattice geometry, with more complicated lattices potentially leading to emergent subsystem symmetries {\it \'{a} la} the mechanism of Ref.~\cite{bulmash2023multipole}. For pursuing the optical lattice experiments mentioned above, it will also be crucial to understand to what extent the physics studied here is modified by the presence of interparticle interactions.

Another intriguing direction is to break the 
time-reversal symmetry of the diffusion process, for example by allowing only ``inward" pair hopping processes while disallowing ``outward'' ones. Such a modification would make the system \emph{active} and is known to result in nonequilibrium phenomena that cannot be captured by entropy maximization~\cite{soto2014run,thompson2011lattice,kourbane2018exact}. 

\paragraph*{Acknowledgments: } We are grateful to Soonwon Choi, Johannes Feldmeier, Byungmin Kang, Andy Lucas, Francisco Machado, Seth Musser, Rahul Nandkishore, Marvin Qi, Brian Skinner, Charles Stahl, Daniel Swartz, and Hongzheng Zhao for helpful discussions and feedback. J.H.H. was supported by the National Research Foundation of Korea(NRF) grant funded by the Korea government(MSIT) (No. 2023R1A2C1002644). He also acknowledges financial support from EPIQS Moore theory centers at MIT and Harvard, where this work was initiated. 

\bibliography{reference}

\bs 

\pagebreak
\newpage 
\bs

	\onecolumngrid
	\begin{center}
		
		\bs 
		\bs 
		\textbf{\large Scaling and localization in multipole-conserving diffusion: \\[.25cm] Supplementary information}\\[.3cm]

	Jung Hoon Han$^1$, Ethan Lake$^{2}$, and Sunghan Ro$^{2}$ \\[.1cm]
	{\itshape ${}^1$Department of Physics, Sungkyunkwan University, Suwon 16419, South Korea\\
		${}^2$Department of Physics, Massachusetts Institute of Technology, Cambridge, Massachusetts 02139, USA} \\
	(Dated: \today)\\[1cm]
\end{center}

\setcounter{equation}{0}
\setcounter{figure}{0}
\setcounter{table}{0}
\setcounter{page}{1}
\renewcommand{\theequation}{S\arabic{equation}}
\renewcommand{\thefigure}{S\arabic{figure}}

\ss*{Contents} 
\begin{itemize}
\item Section \ref{app:quantum_details}: details of calculations relevant for including quantum statistics. 
\item Section \ref{app:cont}: discussion of how to pass from the lattice master equation to the continuum, as well as alternate forms of the master equation which produce modifications to the currents derived in the main text.
\item Section \ref{app:minhop}: analysis of the minimal number of particles needed to participate in multipole conserving hopping processes. 
\item Section \ref{app:langevin}: derivation of the dipolar Langevin equation. 
\item Section \ref{app:crossover}: supplementary plots for the crossover of the dynamic exponent.
\end{itemize}

\bs

\section{Quantum multipolar diffusion}\label{app:quantum_details} 

In this section we provide a few details relating to the incorporation of quantum statistics. We will begin by deriving a master equation which correctly encorporates indistinguishability, and see how this can be used to derive a diffusion equation that holds for any type of statistics (classical, bosonic, or fermionic). We then elaborate on a few phenomenological details of the resulting equilibrium distributions.

\ss{Master equation and diffusion equation}

Our goal in this subsection is to write down an extension of our classical multipole-conserving diffusion process to the setting where the microscopic particles are either bosons or fermions. This can be done with a simple modification of the rates appearing in the master equation, as we now explain. We will mostly focus on the dipole-conserving case, as the extension to higher multipole moments is straightforward. 

\sss{detailed balance} 

We start by demonstrating that the dynamics we consider obeys detailed balance. This follows on general grounds from the microscopic reversibility of the dynamics (see e.g. \cite{crooks2011thermodynamic}), but in the following we provide a more elementary derivation. 

First let us review the case where the statistics is classical. Let $A = \{\r_x^A\}$ and $B = \{\r^B_x\}$ be two configurations of particles related by a dipole hopping process that hops particles at sites $x_1,x_2$ to sites $x_3,x_4$. Let  
$P_{A\ra B}$ denote the probability for a transition between states $A,B$ to occur (and for simplicity assume that $x_1\neq x_2, x_3 \neq x_4$). The most natural generalization of conventional diffusion is to let $P_{A\ra B}$ to be proportional to the number of pairs available to hop, so that  
\be P_{A\ra B} \propto \r^A_{x_1} \r^A_{x_2}, \qq P_{B\ra A} \propto \r^B_{x_3} \r^B_{x_4} = (\r^A_{x_3}+1)(\r^A_{x_4}+1).\ee 
In the main text we stated that this rule satisfies detailed balance. To check this, we need to specify the distribution with respect to which detailed balance is satisfied. This distribution is simply the usual one, which assigns equal weights to each microstate, i.e. each configuration of particles (which for now are treated as being distinguishable). Thus the equilibrium probability of being in state $A$ is 
\be P_A \propto \frac{N!}{\prod_x \r^A_x!},\ee 
where $N$ is the total number of particles. We then see that detailed balance is obeyed, as 
\be \frac{P_A}{P_B} = \frac{\prod_{x=1,2,3,4} \r^B_x!}{\prod_{x=1,2,3,4} \r^A_x!} = \frac{(\rho^A_{x_1}-1)! (\rho^A_{x_2}-1)! (\rho^A_{x_3}+1)! (\r^A_{x_4}+1)!}{\r^A_{x_1}! \r^A_{x_2}! \r^A_{x_3}!\r^A_{x_4}!} = \frac{(\r^A_{x_3}+1)(\r^A_{x_4}+1)}{\r^A_{x_1}\r^A_{x_2}} = \frac{P_{B\ra A}}{P_{A\ra B}}.\ee 

We now generalize to the setting of indistinguishable particles. We claim that the  natural generalization yields transition rates 
\bea P^\z_{A\ra B} &\propto \r^A_{x_1} \r^A_{x_2} (1+\z \r^A_{x_3})(1+\z \r^A_{x_4}),\\ P^\z_{B\ra A} & \propto \r^B_{x_3} \r^B_{x_4} (1+\z \r^B_{x_1})(1+\z \r^B_{x_2}) = (\r^A_{x_3}+1)(\r^A_{x_4}+1)(1+\z\r^A_{x_1}-\z)(1+\z\r^A_{x_2}-\z), \eea 
where $\z = +1,-1,0$ for bosons, fermions, and classical particles, respectively. When $\z=0$ we clearly recover the classical case treated above. When $\z=-1$, the additional factors can be understood simply as a consequence of Pauli blocking; when $\z=+1$ these factors are a consequence of the statistical attraction experienced by bosons. The necessity of including these factors can be seen by checking detailed balance. For $\z=\pm1$, $P^\z_{A\ra B} / P^\z_{B\ra A}$ is equal to unity for all allowed (non-Pauli-blocked) processes. This is consistent with detailed balance because for indistinguishable particles $P_A$ is independent of $A$: the configurations $\{ \r_x\}$ are now themselves microstates, and hence are all assigned equal probabilities in equilibrium. 

\sss{master equation}

With the above modifications to the hopping rates, it is now easy to write down a master equation valid for any value of $\z$. For the 4-site microscopic hopping processes considered in the main text, the full expression is
\bea \label{quantum_master} & \p_t\r  = r (1+\z \r_x) \Big( \r_{x-1} \r_{x+2} (1+\z\r_{x+1}) + \r_{x-2} \r_{x+1} (1+\z \r_{x-1}) \\ & \qq + \r_{x+1} \r_{x+2} (1+\z \r_{x+3}) + \r_{x-2} \r_{x-1} (1+\z_{x+3}) \Big) \\ 
& - r\r_x \Big(\r_{x-1}(1+\z\r_{x+1})(1+\z\r_{x-2}) + \r_{x+1}(1+\z\r_{x-1})(1+\z \r_{x+2}) \\ & \qq  + \r_{x-3} (1+\z \r_{x-2})(1+\z\r_{x-1}) + \r_{x+3}(1+\z\r_{x+1})(1+\z\r_{x+2})\Big) .  \eea 
Master equations for processes conserving higher multipole moments are derived similarly. 

\sss{diffusion equation}

We now perform a derivative expansion on the master equation \eqref{quantum_master}, normalizing the lattice constant to $a=1$. Unilluminating algebra then yields $\p_t \r = -\p_x^2 J$, where the current is now 
\bea \label{quantum_current} J & = D (\r \p_x^2 \r - (\p_x\r)^2 + \z (\r^2 \p_x^2 \r - 2 \r (\p_x\r)^2 ) \\ 
& = D\r^2 (1+\z\r)^2 \p_x^2 \ln \( \frac{1}{1/\r+\z}\)\eea
where $D=4r$ for the 4-site hopping process considered above. 
From the rewriting on the second line, it is easy to derive the steady states 
\be \req(x) = \frac1{e^{\b(x-\mu)} - \z},\ee 
which are the Bose-Einstein ($\z=1$) and Fermi-Dirac ($\z=-1$) distributions that we expect on entropy-maximization grounds. This also suggests that the generalization to $n$-pole conserving dynamics can be done by taking $\p_t = -\p_x^{n+1} J$, where now 
\be J = (-1)^{n+1} D (\r + \z \r^2)^{n+1} \p_x^{n+1} \ln \( \frac1{1/\r + \z}\).\ee 

As a consistency check, we see that regardless of $\z$, our diffusion equation agrees with the $\z=0$ result in the low-density (classical) limit, where the effects of indistinguishability become unimportant. Thus for e.g. dipole-conserving dynamics, the density relaxes with dynamical exponent $z=5$ in all cases. For high-density bosons however, things are different: here the term proportional to $\z$ in \eqref{quantum_current} dominates. Since this term contains three powers of $\r$, the dynamical exponent is instead $z=6$.

\ss{Equilibrium physics}

In this section we elaborate on the phenomenology of the steady states in the bosonic and fermionic cases. 

\sss{bosons} 

As a simple illustrative example, consider $N$ dipole-conserving bosons living in the space $\pp^d \equiv [0,\infty)^d$. For simplicity, we will assume that all $d$ components of the dipole moment $Q^a \equiv \int_\bfr r^a \req(\bfr)$ are equal. A point to note is that, in this section, we will be discussing how the equilibrium distribution changes when the total particle number $N$ is varied. For this reason it will be more convenient to work at fixed dipole moment, rather than at fixed center of mass $r_{\rm cm}^a = Q^a / N$. In this case we may write the equilibrium distribution as 
\be \req(\bfr) = \frac1{e^{\b (\sum_a r_a-\mu)} - 1}\ee  
with $\b>0$. 
The constant $\mu$ --- which by positivity of $\req(\bfr)$ and $\b$ cannot be positive --- is determined by fixing the overall number of particles to be $N$: 
\be \label{nconst} N = \int_\pp  d^dr \frac1{e^{\b(\sum_a r_a - \mu)} - 1}. \ee 
$\b$ on the other hand is determined by fixing the total dipole moment as 
\be Q^a = \int_\pp d^dr \frac{r^a}{e^{\b(\sum_a r_a - \mu)} - 1}. \ee 

Since $\mu$ is monotonically increasing with $N$, there is a ($\b$-dependent) value of $N$ beyond which \eqref{nconst} cannot be satisfied. For larger values of $N$, {\it real-space Bose condensation} will occur, with the remaining bosons occupying the ``zero-energy'' state at $\bfr = 0$. The critical value $N_*$ beyond which a real-space BEC forms is determined by setting $\mu=0$ at fixed $\b$ in \eqref{nconst}, yielding 
\be N_* = \b_*^{-d} \int_\pp d^dr\, \frac1{e^{\sum_a r^a} - 1} \equiv \b_*^{-d} C(d)\ee 
where $\b_*$ will be determined by fixing the dipole moment, and 
\be C(d) = \begin{dcases}
\infty & d =1 \\ 
\pi^2/6 & d=2 \\ 
\z(3) & d= 3 \\ 
\frac{\pi^4}{90} & d=4
\end{dcases}\,\,.\ee 
The divergence for $d=1$ --- which in the present setting is a {\it short-distance} divergence --- prevents a real-space BEC from forming in one dimension. 

We can relate $N_*$ to the dipole moment through the requirement that 
\be Q^a = \b_*^{-(d+1)} \int_\pp d^dr\, \frac{r^a}{e^{\sum_a r^a} - 1} = \b_*^{-(d+1)} C(d+1).\ee 
Solving this equation for $\b_*$ and substituting into the result for $N_*$, we find 
\be N_* =\frac{C(d)}{C(d+1)^{\frac d{d+1}}}  (Q^a)^{\frac d{d+1}} .\ee 
Imagine keeping the dipole moment $Q^a$ fixed while increasing $N$. Initially the ``temperature'' of the Bose-Einstein distribution will decrease with $N$, with the particles becoming progressively more localized around the origin. Once $N > N_*$, all of the remaining particles will be added at the origin, where they join the real-space condensate. From the point of view of the particles away from the origin, particle number conservation is thus spontaneously broken, leading to a phenomenon one might refer to as ``off-diagonal short-range order''. 

Note that since we are fixing the dipole moment --- as opposed to the center of mass --- the number of particles at the origin will increase without bound as $N$ is increased. This can be done at fixed $Q^a$ since adding particles to the origin does not modify the dipole moment (note that this discussion would be modified if we fixed the total center of mass, rather than the total dipole moment). 

Bosons conserving higher multipole moments can be analyzed in much the same way. As an example, consider $N$ quadrupole-conserving bosons living in $\rr^d$. We will again assume for simplicity that all $d$ diagonal components of the quadrupole moment $Q^{aa} = \int_\bfr (r^a)^2 \req(\bfr)$ are equal, and that the off-diagonal components vanish; $Q^{ab} \propto \d^{a,b}$. Following the same analysis as in the dipole case, a real-space BEC forms when the number of particles exceeds the value 
\be N_* = \b_*^{-d/2} \int_{\rr^d}  d^dr\, \frac1{e^{\sum_a (r^a)^2 } - 1} \equiv \b_*^{-d/2} A(d),\ee 
where 
\be A(d) = \begin{dcases} \infty & d=1 \\ \infty & d=2 \\ \pi^{3/2} \z(3/2) & d = 3 \end{dcases}\,\,. \ee 
Here the short-distance divergence which preempts a real-space BEC is stronger than in the dipole-conserving case on account of the ``energy'' $\sum_a (r^a)^2$ vanishing more quickly near the origin (note that the present quadupole-conserving case is formally equivalent to the non-relativistic Bose gas, except with the roles of position space and momentum space interchanged). We may again relate $N_*$ to the (fixed) quadrupole moment $Q^{aa}$ by calculating 
\be Q^{aa} = \b_*^{-(d/2+1)} \int_{\rr^d} d^dr \, \frac{(r^a)^2}{e^{r^2 } - 1} \equiv \b_*^{-(d/2+1)} B(d)\ee 
where 
\be B(d) = \begin{dcases}
\frac{\sqrt\pi}2 \z(3/2) & d=1 \\ \frac{\pi^3}{12} & d=2 \\ \frac{\pi^{3/2} }2\z(5/2) & d=3
\end{dcases} \,\, .\ee 
Therefore we may write the critical particle number in terms of $Q^{aa}$ as 
\be N_* =  \frac{A(d)}{B(d)^{\frac d{d+2}}}(Q^{aa})^{\frac d{d+2}} .\ee 
The interpretation of this equation is similar to the dipole-conserving case: as we increase $N$ past $N_*$ at fixed $Q^{aa}$, the additional particles are all added to the real-space condensate at $\bfr = \bfzero$. 

\sss{Fermions}

\begin{figure}
\includegraphics[width=.35\tw]{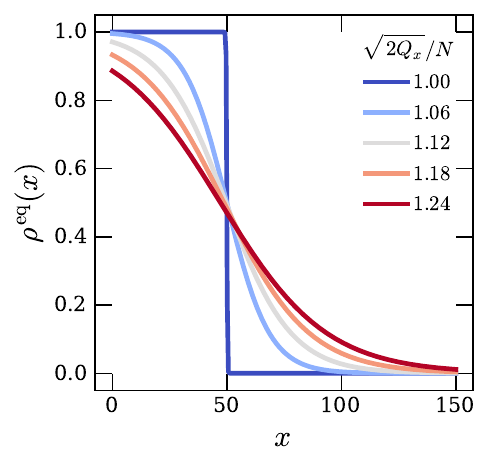}
\caption{\label{fig:fermisurfs} Equilibrium distributions for $N = 50$ dipole-conserving fermions living on the half-space $[0,\infty)$, plotted for various values of the dipole moment $Q_x = \int dx\, x \, \req(x)$. Larger dipole moments yield larger effective temperatures.}
\end{figure}

The analysis for fermions is similar, although of course condensation does not occur. For this reason the physics is similar in all spatial dimensions, and accordingly we will specify to $d=1$ for simplicity. Consider first $N$ dipole-conserving fermions on $\pp = [0,\infty)$. The parameters $\b,\mu$ appearing in the equilibrium distribution are determined through 
\be N = \int_{\pp} dr\, \frac1{e^{\b (r -\mu)}+1},\qq Q^a = \int_{\pp} dr\, \frac{r}{e^{ \b (r -\mu)}+1}.\ee 
The first equation determines $\mu$ as a function of $\b$ via 
\be \b \mu = \ln (e^{\b N} - 1),\ee 
while the second equation implicitly determines $\b$ through 
\be Q^a = - \b^{-2} {\rm Li}_2(1-e^{\b N}).\ee 
The equilibrium distribution for $N=50$ fermions on $\pp$ is shown in Fig. \ref{fig:fermisurfs} for different values of the dipole moment $Q_x$. The minimal value of $Q_x$ is obtained when all of the fermions pile up around the origin; this setup gives $Q_x = N^2/2$. When $Q_x$ is increased the effective ``temperature'' of $\req(x)$ also increases, as seen in the figure. Higher multipole moments are analyzed similarly.

\section{continuum limits, scaling, and modified master equations}\label{app:cont}

In this appendix we discuss different types of master equations and the scaling relations needed to obtain continuum differential equations from them. We will restrict our attention to one-dimensional dipole-conserving systems for concreteness; more general settings can be analyzed similarly. 

\ss{Continuum limits}

As a review, first consider how one would derive the {\it conventional} diffusion equation from a microscopic master equation. The simplest setting involves non-interacting classical particles hopping in 1d. Here one considers a process such that in each time interval $dt$, any given particle has a probability $r\,dt$ of hopping one lattice site to the left, and the same probability of hopping one site to the right. On a site $x$ with $\r_x$ particles, the probabilities that {\it some} particle hops to the left or the right are consequently equal to $r\, dt \r_x$. Summing up the contributions from particles which hop into and out of a given site $x$, we then have 
\be \r_x(t+dt) = \r_x(t) + r\,dt(\r_{x+1}(t) + \r_{x-1}(t) - 2 \r_x(t)),\ee 
which in the limit $dt\ra 0$ gives the master equation
\be \p_t \r_x = r (\r_{x+1} + \r_{x-1} - 2\r_x).\ee 
One scaling limit which produces a continuum differential equation is as follows. Consider a system containing $L$ lattice sites with lattice spacing $a$. We then take $a \ra 0$ while sending $r \ra \infty, L\ra \infty$ in such a way that $ra^2$ and the dimensionful system size $aL$ remain fixed. This then produces the usual diffusion equation 
\be \p_t \r = D \(\p_x^2 \r + \frac{a^2}{12} \p_x^4 \r + \cdots\),\qq D \equiv ra^2 ,\ee 
where the $\cdots$ contain higher derivatives and higher powers of $a$. In the usual continuum limit we imagine that all of the continuum derivatives of $\r_x$ are of the same order, and hence after sending $a\ra0$ only the first term survives.

In order to have a well-defined continuum limit, we require that $\r_x$ be finite, and not parametrically divergent or vanishing as $a \ra 0, r,L\ra \infty$. In terms of the original lattice model, this means that we must consider the {\it dilute limit} where $n_i \ra 0$. In other words, this continuum limit corresponds to keeping the number of particles present in the system fixed while simultaneously fine-graining the lattice. 

This type of continuum limit --- one which corresponds to the dilute limit of a lattice model with lattice spacing $a\ra0$ --- is not the correct one to take for the kinematically constrained systems studied in this paper. Indeed, if we let the number of lattice sites over which any given elementary hopping process acts be finite, the kinetic constraints then completely prevent all motion in this limit, and the particles all become trivially localized. We can get around this by simultaneously taking the range of the elementary particle hopping processes to diverge as $1/a$, but this produces a system (a dilute multipole-conserving continuum gas) whose dynamics is potentially very different from the non-dilute lattice gases studied in the present work. 

Instead, the limit we are interested in is one where the average number of particles on each lattice site is nonzero, and for typical configurations is larger than the critical value below which the dynamics begins to freeze out (as mentioned above, this critical value is set by the range of the elementary hopping processes \cite{skinner23}). For $\r_x$ to be well-defined we thus need to keep the lattice spacing {\it fixed}, with a derivative expansion of $\r_x$ being performed simply under the assumption that near equilibrium, high-order derivatives of $\r_x$ are numerically smaller than low-order ones, but not parametrically smaller in any scaling parameter. For example, in the case of particles living on the half-line $x >0$ we have (4) of the main text%
\be  \label{reqcmapp} \req(x) = \frac N{x_{\rm cm}}e^{-x/x_{\rm cm}},\ee and higher derivatives are thus suppressed in powers of $a/x_{\rm cm}$, which is finite (but still usually rather small) in the limit we are interested in.  

We note that these sorts of subtleties, namely the existence of multiple inequivalent ways of taking the continuum limit, have also recently been discussed from a field theoretic point of view in e.g. Ref. \cite{gorantla2021low}. 

\ss{Different forms of the master equation}

We now take a critical look at the 1d dipolar master equation, and show that it can take on qualitatively different forms depending on the type of elementary hopping process considered. This analysis will also help further motivate the preceeding discussion of continuum limits and will help sharpen up the meaning of ``high density regime''. 

As discussed in the main text, the 4-site elementary hopping process ``$0110\lra 1001$'' captured by (5) and (6) gives rise to the master equation 
\be 
\partial_t \r_x = r (\r_{x-1} \r_{x+2} + \r_{x-2} \r_{x+1} + \r_{x+1}\r_{x+2} +\r_{x-2} \r_{x-1})  - r \r_x (\r_{x-1} + \r_{x+1} + \r_{x-3} + \r_{x+3}), \ee 
where $r$ is again the hopping rate. 
Upon performing a derivative expansion of $\r$ as in the main text, we obtain the dipolar diffusion equation $\p_t \r = - \p_x^2 J_{4-{\rm site}}$ with current 
\be J_{4-{\rm site}}= 4ra^5( (\p_x\r)^2 - \r \p_x^2 \r  + a^2 (\cdots)),\ee 
where the $\cdots$ are suppressed by higher powers of derivatives. 

It is instructive to consider modifying the form of the elementary dipole-conserving hopping processes that go into the derivation of the master equation. The most interesting modifications are those where two particles leave from or arrive at the {\it same} site (other modifications simply lead to different values for $D$). For example, consider the 3-site ``$020\lra 101$'' hopping process, which is a more compact version of the 4-site process considered above. The change in particle occupancies for a hopping process occuring on sites $x,x+1,x+2$ is 
\be (\De \r_x , \De \r_{x+1}, \De \r_{x+2}) = (\pm1,\mp2,\pm1)\ee 
with the top sign corresponding to an ``out'' hopping process and the bottom to an ``in'' process. Within the same type of mean-field approximation used in the main text, the rate for the ``in'' process is $r dt \r_x \r_{x+2}$, while the rate for the ``out'' process is proportional to the number of pairs on site $i$, viz. $rdt \r_{x+1} (\r_{x+1}-1)$ (it is easy to verify that these rates satisfy detailed balance). 

The important part here is that the rate for the ``out'' process contains a term {\it linear} in $n$. This produces a master equation 
\be \p_t \r_x = r(2\r_{x-1}\r_{x+1} + \r_{x+1}(\r_{x+1}-1) + \r_{x-1}(\r_{x-1} - 1) - \r_x(2(\r_x-1) + \r_{x+2} + \r_{x-2}))
\ee 
which contains terms linear in $n$. 
Performing a derivative expansion then yields $\p_t \r = - \p_x^2 J_{3-{\rm site}}$, with the current now given by 
\be  \label{threej} J_{3-{\rm site}} =  ra^5 ((\p_x\r)^2 - \r \p_x^2 \r + a^2 (\cdots) - a^{-3} \r ),\ee 
where the $\cdots$ are terms quadratic in $\r$ that contain four derivatives. This expression for the current differs from the 4-site version by the extra $- a^{-3} \r$ term on the RHS, which represents a diffusive term with a {\it negative} effective diffusion constant, favoring charge localization (interestingly, general hydrodynamic arguments can be given which would have appeared to rule out the existence of such a term \cite{guo2022fracton}). 

Dropping the terms in $(\cdots)$, the zero-current solutions obtained from \eqref{threej} are no longer simple exponentials, and instead take the form
\be \req_{3-{\rm site}}(x) = \frac{a^{-3}}{2\a^2} \sinh^2(\a x + \b),\ee 
where $\a,\b,$ are fixed by the total number of particles and total dipole moment. 

Note that if we were to take the standard continuum limit (in which we send $a \ra 0$ while keeping $\r$ finite), the last $a^{-3} \r$ term in $J_{3-{\rm site}}$ would dominate, and the physics would look very different to the 4-site variant. 
Despite this, we emperically find that simulations run with only 3-site hoppings give results essentially identical to those run with only 4-site hoppings, at least for the large-density regimes studied in the main text, where the dynamics is more universal (in both cases the steady-state distributions are exponentially localized, and the dynamic exponent is $z=5$). Accordingly, these simulations must be run in a regime where the $(\p_x \r)^2 , \r \p_x^2\r$ terms are not subleading compared to $a^{-3}\r$. This necessitates that the average density be larger than the amount by which higher derivatives of $\r$ are suppressed. For the distribution with characteristic length scale $\ell$ (e.g. $\ell = x_{\rm cm}$ for \eqref{reqcmapp}, or $\ell = \s$ for the Gaussian $\r(x) = \frac{N}{\sqrt\twp\s}e^{-x^2/2\s^2} $, the $(\p_x \r)^2 , \r \p_x^2\r$ terms dominate when 
\be N \gtrsim (\ell / a)^3,\ee 
which is satisfied in the ``high-density'' situations considered in the main text. For now we will content ourselves with this characterization, but in the future it will be valuable to precisely characterize the universal features (or lack thereof) exhibited by general multipolar dynamics in different parameter regimes.

\section{minimal nonlinearities of the consitutive relation} \label{app:minhop}

In this section we lower-bound the minimal number $m$ of powers of $\r$ that must enter into the constitutive relation for the current in $n$-pole conserving diffusive dynamics, restricting to 1d lattice systems for simplicity. In the main text we saw that the current was quadratic in $\r$ ($m=2$) in the dipole-conserving case ($n=1$), and cubic in $\r$ ($m=3$) in the quadrupole-conserving case ($n=2$) \footnote{The nonlinearity measured by $m$ discussed here refers to the largest power of the density appearing on the RHS of the master equation. When the hopping process is such that at most one particle hops from any given site, all terms on the RHS will be of the same order. In other cases (such as the 3-site dipole hopping process discussed in App. \ref{app:cont}) smaller powers may additionally appear. In this appendix we are interested in determining the minimal value that the largest power on the RHS can be.}.
By physicist's induction, it is then tempting to conclude that $m=n+1$. While this (likely) is actually not true in complete generality, we are able to prove the following proposition, which show that $m\geq n+1$: 

\bs 

\noindent {\bf Proposition. } Consider particles moving on a 1d lattice under translation-invariant dynamic that conserves the $n$-pole moment of the charge. Then the constitutive relation expressing the current in terms of the density $\r$ must involve at least $m \geq n+1$ powers of $\r$, and at least $n+1$ spatial derivatives. This lower bound can be saturated for at least all $n \leq 10$, as well for $n=12$. 

\bs 		

{\it Proof:} 
The number $m$ is determined by finding the smallest number of particles that can participate in an $n$-pole preserving particle hopping process. For a process involving $m$ particles, let $\bfa,\bfb \in \zz^m$ denote the initial and final positions of the particles, respectively. Note that if $a_i = b_j$ for any $i,j \in 1,\dots,m$, then at least one particle does not actually move during the hopping process; thus wolog we may restrict to $\{ a_i \} \cap \{ b_i \} = \emptyset$. 

For translation-invariant dynamics, the conservation of the $n$th multipole moment implies the conservation of the $p$-pole moments for all $p<m$. Thus since the process $\bfa \mt \bfb$ must preserve all multipole moments up to $n$, we require that 
\be \label{symsum} \sum_i a_i^p = \sum_i b_i^p \, \, \forall \, \, p \in \{0,\dots,n\}.\ee 
The goal is thus to find the smallest-length pair of disjoint integer-valued sets such that the above equation is satisfied. 

We now show that \eqref{symsum} can be satisfied only when $m>n$. Indeed, suppose that $m = n$. It is a well-known fact \cite{macdonald1998symmetric} that the set of ``power-sum polynomials''
\be P_p(\bfa)\equiv \sum_{i=1}^m a_i^p\ee 
for $p\in \{0,\dots,m\}$ generate all symmetric polynomials in $m$ variables. This means that the symmetric polynomials $\prod_{i=1}^m a_i, \prod_{i=1}^m b_i$ can be expressed as linear combinations of the $P_p(\bfa), P_p(\bfb)$, respectively. But since we have assumed that all $o\leq m$ multipole moments are conserved, we have 
\be P_p(\bfa)  = P_p(\bfb)\, \, \forall \,\, p \in \{0,\dots,m\} \, \, \implies \, \, \prod_{i=1}^m a_i = \prod_{i=1}^m b_i.\ee 

Now we use the fact that \eqref{symsum} is invariant under translations, in that \eqref{symsum} implies 
\be P_p(\bfa-x) = P_p(\bfb-x) \ee 
for any constant vector $x$ (this follows from the fact that $P_p(\bfa -x)$ is also a symmetric polynomial). In particular, we may choose $x = a_1$ to set $a_1 = 0$ wolog. But then from the above we see that since $\prod_i a_i = 0$, we must also have $\prod_i b_i = 0$. In particular, one of the $b_i$ must equal zero, which contradicts our assumption about the disjointness of $\{a_i\}, \{b_i\}$. 

This shows that any $n$-pole conserving hopping process must involve at least $n+1$ particles, so that the constitutive relation must involve at least $n+1$ powers of the density. This also implies that the current must involve at least $n+1$ derivatives; if it involved fewer than $n+1$ derivatives then it would be proportional to an overall power of the density $J = \r^k( \cdots)$ (where $\cdots$ involves derivatives of $\r$), and the $\r^k$ in front could be stripped away without affecting the nature of the zero-current solutions. 

We have shown that $n+1$-body terms are necessary, but not that they are sufficient. Showing this would amount to proving that one can always find two mutually disjoint size-$n+1$ sets of integers satisfying \eqref{symsum}, which is in fact a famous open problem in number theory known as the Prouhet-Tarry-Escott problem \cite{borwein2002computational}. We will of course not attempt to make progress on this problem here, and simply note that solutions are known to exist for $n\leq 10$ and $n=12$, but that for other values of $n$ the answer is unknown (with the existence of solutions at large $n$ seeming unlikely) \cite{borwein2002computational}. \qed 

\bs 

The explicit $m=n+1$ solutions of \eqref{symsum} (when they exist) quickly get rather complicated for large $n$, meaning that they involve very long-range hopping processes. For example, when $n=7$ the simplest solution is an eight-body hopping process with 
\bea a & = [1, 5, 10, 24, 28, 42, 47, 51]\\ 
b & = [2, 3, 12, 21, 31, 40, 49, 50]\eea 
meaning that the minimal-body hopping process extends over 50 lattice sites. 

Depending on the physical context it may be unreasonable to consider such long-range hopping terms, instead considering shorter-range but higher-body hopping processes. The minimal-range hopping processes involve a number of particles that scales exponentially in $n$ but has range equal to $n+2$, which can be constructed with the aid of Pascal's triangle by letting 
\be \bfa = \bigoplus_{i=0}^{\lfloor (n+1)/2 \rfloor} (2i)^{\oplus {n+1 \choose 2i}},\qq \bfb = \bigoplus_{i=1}^{\lfloor (n+2)/2\rfloor} (2i-1)^{\oplus {n+1\choose 2i-1}}, \ee 
where $\oplus$ denotes concatenation, so that e.g. $x^{\oplus k}$ is the length-$k$ vector with all entries equal to $x$. For dipoles this gives the 2-body process $\bfa = (1,3), \, \bfb = (2,2)$, for quadrupoles the 4-body process $\bfa = (1,3,3,
3), \, \bfb = (2,2,2,4)$, and so on. Processes involving a non-minimal number of particles can be checked to lead to difussion equations with additional powers of $\r$ appearing in the constitutive relation for the current. For example, the (non-minimal) 4-body quadrupole-conserving process just described gives rise to the relation $J = D \r^4 \p_x^3 \ln(\r)$, which has one more power of $\r$ as compared to the minimal solution $J = D \r^3 \p_x^3 \ln (\r)$.

\section{Derivation of the dipolar Langevin equation} \label{app:langevin} 

The dipolar diffusion equation given in the main text is deterministic, and misses the stochastic nature of the pairwise hopping. Properly including fluctuation effects can be done by statistically analyzing the full microscopic motion and coarse-graining the result, as laid out e.g. in Ref.~\cite{lefevre07}. We will use notation appropriate for a 1d chain; generalizations to higher dimensions are straightforward. We will work in units where the lattice spacing $a=1$ throughout. 

In each time step $t \to t+dt$, each pair of particles in the system undergoes pairwise diffusion with a rate $r$, causing the number of particles in each lattice to change by $\Delta \r_x(t) \equiv \r_x(t+dt) - \r_x(t)$. Then $\Delta \r_x(t)$ for all possible $x$ and $t$ together determines the trajectory of the system.

Next, we introduce the moment-generating function of the particle dynamics
\begin{equation}
W[\hat\r_x(t)] = \Big\langle \exp \Big[ \sum_{x,t} \hat\r_x(t) \Delta \r_x(t)  \Big]
\Big\rangle , \label{eq:generating-function} 
\end{equation}
where $\hat\r_x(t)$ is the conjugate field of $\Delta \r_x(t)$ and the average is taken over all possible trajectories of $\Delta \r_x(t)$. In the continuum time limit, functional derivatives of $\ln W$ with respect to $\hat \r$ can be used to calculate connected correlation functions of $\p_t \r$. 

Since the pairwise diffusive jump processes are independent, we can rewrite the generating function as 
\begin{equation} \label{eq:w_n}
W[\hat\r_x(t)] = \prod_{x,t} \Big\langle \exp \Big[\hat\r_x(t) \Delta \r_x(t)  \Big]
\Big\rangle .
\end{equation}
For a lattice with $L$ sites and periodic boundary conditions, there are $2L$ possible jump processes; the {\it ingoing} ($1001 \to 0110$) and {\it outgoing} ($0110 \to 1001$) process for any given adjacent quartet of sites $(x, x+1, x+2, x+3)$.

To calculate Eq.~\eqref{eq:w_n} over these $2L$ processes, we refer to $(x,x+1,x+2,x+3)$ sites with the index $i$  and take an average over ingoing and outgoing processes. To make the subsequent expressions concise, we omit $t$ from the arguments of $\r_x(t)$ and $\hat\r_x(t)$ in the following. We obtain 

\begin{align}
\nonumber W[\hat\r_x(t)] &= \prod_{x,t} \Big[   \r_{x+1} \r_{x+2} r dt e^{ \hat\r_x - \hat\r_{x+1} - \hat\r_{x+2} + \hat\r_{x+3} } + \r_{x}\r_{x+3} r dt e^{ -\hat\r_x + \hat\r_{x+1} + \hat\r_{x+2} - \hat\r_{x+3} } \\
\label{eq:w_n_long} & ~~~~~~~~~~~~~  + 1 - [ \r_{x+1} \r_{x+2} + \r_{x}\r_{x+3} ]  r dt \Big].
\end{align}

In the square brackets of the expression above, the first term is the product between the probability for the outgoing motion to occur, $\r_{x} \r_{x+3} r dt$, and the conjugate field factor associated with the changes of the particle numbers $\Delta \r_{x} = \Delta \r_{x+3} =  +1$ and $\Delta \r_{x+1} = \Delta \r_{x+2} = -1$. Similarly, the second term consists of the probability of the ingoing motion and the conjugate field factor with $\Delta \r_{x} = \Delta \r_{x+3} = -1$ and $\Delta \r_{x+1} = \Delta \r_{x+2} = +1$. The last term is the probability for no motion to occur, with the corresponding conjugate factor with $\Delta \r_{x} = \cdots = \Delta \r_{x+3} = 0$. Now, we introduce
\bea 
R_i^+ &\equiv r \r_{x+1} \r_{x+2} \,, \\
R_i^- &\equiv r \r_{x} \r_{x+3} \,, \\
S_i &\equiv \hat\r_x - \hat\r_{x+1} - \hat\r_{x+2} + \hat\r_{x+3}\,
\eea
and write Eq.~\eqref{eq:w_n_long} as
\bea 	\label{eq:w_n_abb}
W[\hat\r_x(t)] &= \prod_{i,t} \Big[ 1 + dt \big\{ R_i^+ (e^{S_i}-1) + R_i^- (e^{-S_i}-1) \big\} \Big], \\
&\simeq \exp \Big[ \sum_{i,s} dt \{  R_i^+ (e^{S_i}-1) + R_i^- (e^{-S_i}-1) \}  \Big].
\eea 
To proceed, we take the continuum time limit and expand $\hat \r$ in gradients (which is allowed as long as we are only interested in long-distance correlation functions of $\De \r$), allowing us to write $S_x \approx 2\p_x^2\r$. This gives 

\bea \label{lnw}  \ln W[\hat\r_x(t)] &= \sum_x \int_t ( R^+_x (e^{2\p_x^2 \hat\r} - 1) + R^-_x (e^{-2\p_x^2 \hat \r}-1) ).
\eea 

Taking the variation with respect to $\hat{\rho}$ and then sending $\hat{\rho} \rightarrow 0$ gives the average $\langle \partial_t \rho \rangle$, yielding 
\bea \langle \partial_t \rho \rangle &  = 2\p_x^2 (R^+_x - R^-_x) \\ &
\ra 
D \partial_x^2  \left[ (\partial_x \rho )^2 - \rho \partial_x^2 \rho  \right] , 
\eea 
where in the second line we have defined $D \equiv 4r$ and expanded $\r_x$ up to quadratic order in derivatives. We thus recover the dipolar diffusion equation, but now with $\partial_t \rho$  replaced by the statistical average $\langle \partial_t \rho \rangle$. 

Taking two functional derivatives of \eqref{lnw} gives 
\bea	\label{eq:second-cumulant}  \langle  \partial_t \rho (x, t) \partial_{t'} \rho (x', t') \rangle_c 
& =  4 \p_x^2 \p_{x'}^2 [ \d(x-x')\d(t-t')(R^+_x + R^-_x)] \\ 
& \ra 	
2 D \partial_x^2 \partial_{x'}^2 \left[  \rho^2 (x,t) \delta (x - x') \delta (t - t') \right] . 
\eea 
In order to reproduce this result, one must supplement the dipolar diffusion equation with the fluctuation term, which results in the dipolar Langevin equation (10) quoted in the main text, 
\be \partial_t \rho (x, t) = -D\partial_x^2 (\r^2\p_x^2 \ln(\r)) + \sqrt{2D} \p_x^2  \eta (x, t),
\ee 
where the noise field $\eta$ obeys $\langle \eta (x, t) \eta(x', t') \rangle = \delta (t - t')  \delta (x - x') \r^2(x,t)$.

One application of this is a derivation of the structure factor governing the behavior of small density fluctuations about equilibirum. To this end, write $\r = \req + \d\r$, and linearize the Langevin equation in $\d\r$ while dropping all gradients of $\req$ (this amounts to considering fluctuations on length scales short enough so that $\req$ can be treated as being approximately constant). Then 
\be \p_t \d\r = - \wt D  \p_x^4 \d\r + \sqrt{2D} \p_x^2 \eta,\ee 
where $\wt D \equiv D (\req)^2$ is an effective subdiffusion constant, and $\lan \eta(x,t) \eta(x',t') \ran_c = (\req)^2\d(t-t') \d(x-x')$. Going to Fourier space gives 
\be \lan \d\r(k,t) \d\r(-k,t) \ran_c = \int \frac{d\o}\twp \frac{2 \wt D \req k^4}{\o^2 + (\wt D k^4)^2 } = \req,\ee 
which is the exact same result one would obtain by applying this formalism to the regular diffusion equation.

\section{crossover of the dynamic exponent} \label{app:crossover}

\begin{figure} 
\includegraphics[width=1\tw]{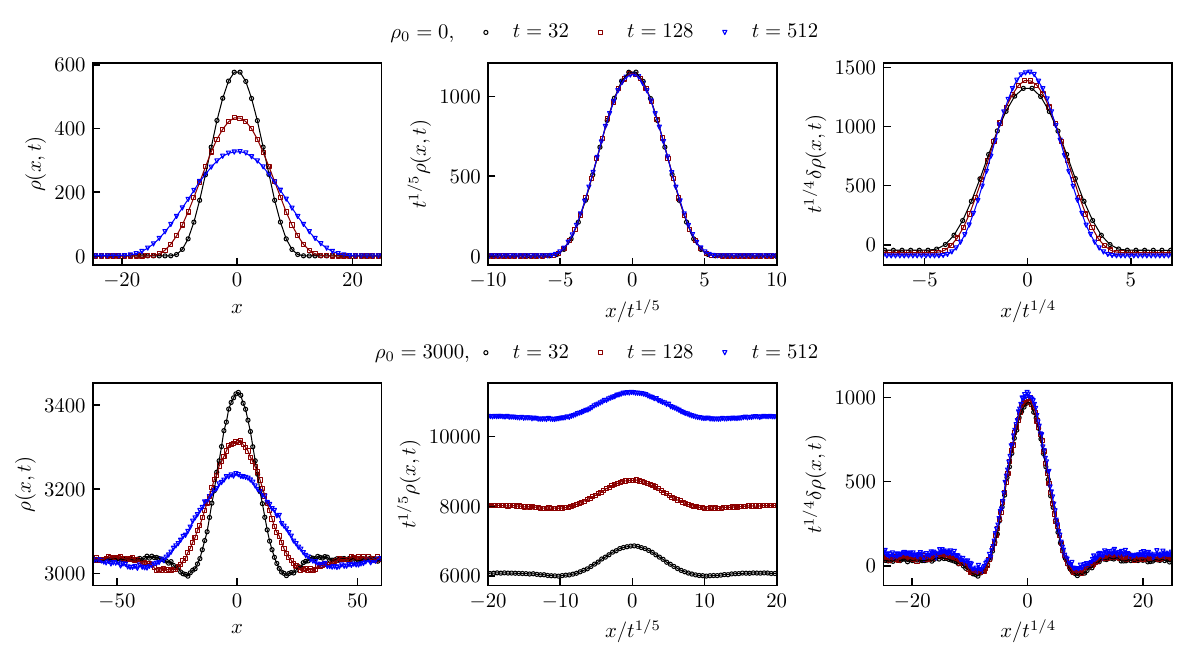} 
\caption{\label{fig:scaling} Direct demonstration of dynamic exponent crossover. The upper and lower panels show the time evolution of the density field of $N_0 = 6000$ particles initially positioned at $x=0$, with background densities $\rho_0 = 0$ (upper) and $\rho_0 = 3000$ (lower) respectively. The left panels show the bare density profiles. The center panels present the density profiles after scaling using $z=5$, while the right panels show scaling results of the density difference field $\delta \rho(x,t)$ with $z=4$. 
}\label{Fig:SM_scaling} 
\end{figure}

In this section, we numerically demonstrate the crossover of the dynamic exponent by directly studying the scaling of the density and the density difference profiles for dipole-conserving diffusion. In one dimension, our study suggests that the density field $\rho(x,t)$ evolving in a zero-density background satisfies dynamic scaling with $z = 5$. This means that, in the long-time limit, density profiles at different times become asymptotically self-similar to each other when the spatial coordinate is scaled as $x/t^{1/z}$, and the density field is scaled as $t^{1/5} \rho(x,t)$. 
On the other hand, we expect that the density difference $\delta \rho(x,t) = \rho(x,t) - {\rho}_\mathrm{eq}(x)$ relative to the equilibrium density profile will adhere to the scaling with a dynamic exponent of $z = 4$ when the density difference is small compared to the equilibrium density. 

This is directly demonstrated in Fig.~\ref{Fig:SM_scaling} through the analysis of density profiles of 6,000 particles evolving in constant density backgrounds with the densities $\rho_0 = 0$ and $3000$. For particles evolving in the zero-density background, as shown in the upper panels, the density profile $\rho(x,t)$ shows an excellent collapse upon scaling with $z=5$. However, the density difference fields $\delta \rho(x,t)$ obtained in different times scaled with $z=4$ do not align with each other. 
In contrast, the particles dispersing in a high-density background, displayed in the lower panels, exhibit a collapse in their density difference field when scaled with $z=4$. This result indeed confirms the crossover from $z=5$ to $z=4$ as the density difference becomes comparable to the equilibrium density.

\bibliography{reference}

\bs 

\pagebreak
\newpage 
\bs 

\end{document}